# Boosting LLM-based Relevance Modeling with Distribution-Aware Robust Learning

Hong Liu*
yizhou.lh@antgroup.com
Ant Group
Hangzhou, China

Saisai Gong*
gongsaisai.gss@antgroup.com
Ant Group
Hangzhou, China

Yixin Ji
jyx402906@antgroup.com
Ant Group
Hangzhou, China

Kaixin Wu
daniel.wkx@antgroup.com
Ant Group
Hangzhou, China

Jia Xu†
steve.xuj@antgroup.com
Ant Group
Hangzhou, China

Jinjie Gu
jinjie.gujj@antgroup.com
Ant Group
Hangzhou, China

## ABSTRACT

Relevance modeling plays a crucial role in e-commerce search engines, striving to identify the utmost pertinent items corresponding to a given search query. With the rapid advancement of pre-trained large language models (LLMs), recent endeavors have leveraged the capabilities of LLMs in relevance modeling, resulting in enhanced performance. This is usually done through the process of fine-tuning LLMs on specifically annotated datasets to determine the relevance between queries and items. However, there are two limitations when LLMs are naively employed for relevance modeling through fine-tuning and inference. First, it is not inherently efficient for performing nuanced tasks beyond simple yes or no answers, such as assessing search relevance. It may therefore tend to be overconfident and struggle to distinguish fine-grained degrees of relevance (e.g., strong relevance, weak relevance, irrelevance) used in search engines. Second, it exhibits significant performance degradation when confronted with data distribution shift in real-world scenarios. In this paper, we propose a novel Distribution-Aware Robust Learning framework (DaRL) for relevance modeling in Alipay Search. Specifically, we design an effective loss function to enhance the discriminability of LLM-based relevance modeling across various fine-grained degrees of query-item relevance. To improve the generalizability of LLM-based relevance modeling, we first propose the Distribution-Aware Sample Augmentation (DASA) module. This module utilizes out-of-distribution (OOD) detection techniques to actively select appropriate samples that are not well covered by the original training set for model fine-tuning. Furthermore, we adopt a multi-stage fine-tuning strategy to simultaneously improve in-distribution (ID) and OOD performance, bridging the performance gap between them. DaRL has been deployed online to serve the Alipay's insurance product search. Both offline experiments on real-world industry data and online A/B testing show that DaRL effectively improves the performance of relevance modeling.

## 1 INTRODUCTION

E-commerce search engines, like Alipay Search [1], enable users to quickly and efficiently find desirable products, services, or other types of items that meet their information needs. Relevance modeling holds paramount importance in e-commerce search engines, which prioritizes the most relevant item with the query to enhance user experience.

Over the past several decades, a large number of methods have been developed to model search relevance, transitioning from traditional heuristic and learning-based models [11, 16, 32] to more effective deep learning models [10, 15, 41]. In recent years, the advent of pre-trained language models like BERT [5] has revolutionized best practices in deep learning-based relevance modeling. A variety of studies have utilized BERT and other variations of pre-trained language models to capture the intricate contextual semantics of queries and items and to encode the semantic relatedness between them [23, 33, 47], leading to a significant improvement in relevance modeling. More recently, the rise of LLMs like GPT [2], LLaMA [38], GLM [7] has marked an important milestone in natural language processing (NLP). With extensive world knowledge encoded in model weights and inherent strong reasoning abilities, LLMs have achieved state-of-the-art performance on multiple NLP tasks [26]. The remarkable capability of LLMs in textual understanding has motivated several studies to harness the power of LLMs in relevance modeling or judgment [3, 8, 24, 25, 30, 37]. These methods leverage pre-trained LLMs as the backbone of modeling, and then fine-tune LLMs on task-specific annotated datasets to determine query-item relevance, achieving promising performance improvements.

---

*Both authors contributed equally to this research.
†Corresponding author.

[1]https://www.alipay.com/

Although fine-tuning empowers us to tailor LLMs to excel in relevance modeling and achieve superior performance, naive fine-tuning of LLMs suffers from two major limitations. First, it is not inherently well-suited for nuanced tasks like assessing search relevance. To differentiate between different query-item relevance and enhance user experience, e-commerce search typically requires multiple fine-grained relevance degrees, such as strong relevance, weak relevance, or irrelevance, beyond simple yes or no answers. Discerning fine-grained relevance degrees in e-commerce search could be a nuanced and subjective task due to industry-specific lingo and nuances. While LLMs have excellent understanding of general world knowledge, they are not inherently equipped to fully grasp the nuances of specialized relevance modeling tasks. As a result, without adapted fine-tuning methods, vanilla fine-tuned LLMs struggle to distinguish fine-grained degrees of relevance and tend to provide biased or overconfident relevance estimation [40]. Second, naive fine-tuning of LLMs showcases significant performance degradation when facing with data distribution shift in real-world applications [6]. Thanks to large-scale pre-training on massive text corpus by following instructions and aligning with human preference [28], pre-trained LLMs exhibit impressive generalization capabilities. However, naively fine-tuning pre-trained LLMs to a specific task can lead to model over-specialization to the fine-tuning data [39], which largely hinders LLMs' intrinsic generalization ability on real-world OOD data. As a result, there exists a gap between the performance of vanilla fine-tuned LLMs on ID and OOD samples.

To address the aforementioned issues, we propose the DaRL framework that leverages data-efficient robust fine-tuning coupled with prediction difference regularization to improve the generalizability and discriminability of LLM-based relevance modeling. Specifically, we grade LLM outputs for relevance prediction, and enhance the model discriminability of fine-grained query-item relevance degrees by regularizing the prediction difference of distinct relevance degrees based on Kullback-Leibler (KL) divergence. To improve the generalizability of LLM-based relevance modeling, we propose the DASA module, which uses OOD detection techniques to actively gather new OOD samples that are not well covered in the original training set for fine-tuning. DASA effectively enriches the fine-tuning dataset and provides the fine-tuned LLMs with more accurate and relevant insights for OOD generalization. Furthermore, we employ a multi-stage fine-tuning strategy to simultaneously improve the ID and OOD performance, wherein we first learn a near-optimal final layer, then fine-tune the entire model, and finally conduct model weight interpolation. In addition, we also enhance the discriminability of relevance modeling through the application of prediction difference regularization techniques.

Our main contributions are summarized as follows:
- To the best of our knowledge, we are the first to explore the synergy of distribution-aware data enrichment and robust fine-tuning strategies, for improving both the generalizability and discriminability of LLM-based relevance modeling.

- We propose DaRL with three main components. First, DASA leverages OOD detection techniques to effectively enhance the model's generalizability from a data enrichment perspective. Second, a multi-stage fine-tuning strategy is proposed to maintain ID performance while further improving OOD generalizability. Third, we introduce an auxiliary loss based on KL divergence to reduce model overconfidence and strengthen discriminability.
- We demonstrate the effectiveness of DaRL through experiments on real-world industrial datasets. It has been deployed in our online e-commerce search engine, resulting in a significant improvement in user search experience.

## 2 RELATED WORK

In the realm of relevance modeling, early works mainly relied on feature engineering, such as lexical [32], syntactic [1], semantic [11] and user behaviour features [16], to measure the relevance between a query and an item. These traditional heuristic or learning-based methods usually have limitations like vocabulary mismatch and lack of deep semantic understanding [9]. With the development of deep neural networks, modern relevance modeling has shifted to more effective deep learning methods [10, 15, 41], which can obviate the need for laboriously hand-crafted features and overcome the limitations of traditional approaches.

Based on whether query-item interaction is captured, deep learning based relevance modeling can be broadly categorized into representation-based methods and interaction-based ones. For example, DSSM [15] and C-DSSM [35] are two representation-based methods that encode query and item independently without interaction, while DRMM [10] and MatchPyramid [29] are representation-based ones that capture the rich interaction structures between query and item from token or phrase level. As a result, interaction-based approaches are usually more expensive but more effective than representation-based ones.

In the past few years, pre-trained transformer-based language models, such as BERT [5] and RoBERTa [21], have shown their superiority in relevance modeling and are widely used as the backbone of deep modeling. Relevance modeling based on BERT or its counterparts can effectively capture the intricate contextual semantics of query and item, thereby significantly enhancing the modeling performance. Sentence-BERT [31] uses siamese and triplet network structures to derive sentence embeddings, facilitating semantic similarity search. TwinBERT [23] is a representation-based model that enhances its performance by knowledge distillation. Hofstätter et al. [13] combined a topic-aware batch sampling with a dual-teacher supervision to enhance the representation-based BERT dense retriever. ColBERTv2 [33] introduces late interaction between the query and item for improved performance.

In recent times, LLMs, such as GPT-4 [27], LLaMA [38] and GLM [7], have marked a breakthrough in NLP and demonstrated strong capabilities in language understanding, generating and reasoning. A few works have leveraged the power of LLMs for relevance modeling in search engines, achieving a promising improvement. Chen et al. [3] combined a fine-tuned LLM with the behavior-based relevance model to enhance the query-item matching performance. Ma et al. [24] fine-tuned LLaMA for multi-stage text retrieval. Peng et al. [30] developed a series of e-commerce LLMs called eCeLLM by instruction-tuning general-purpose LLMs on a high-quality benchmark instruction dataset.

Fine-tuning is usually performed by reformulating the downstream task as a language modeling problem using prompts or templates [28, 30, 34], with the goal of optimizing the prediction of target tokens or sequences. To avoid the high computational costs of full fine-tuning, numerous Parameter-Efficient Fine-Tuning (PEFT) methods have been proposed to achieve acceptable fine-tuning performance at reduced costs, such as LoRA [14], prompt tuning [20] and adapters [12]. Although pre-trained LLMs exhibit impressive generalization ability across tasks, fine-tuning may impair the LLMs' intrinsic generalization ability due to shortcut learning [6], feature distortion [18] and other reasons, leading to significant performance degradation when confronted with distribution shift in real-world scenarios [42, 44, 45]. Adapted from linear probing and then full fine-tuning (LPFT) strategy [18], EH-FT [44] first performs head pre-training using PEFT and then conducts full fine-tuning to improve the ID generalization capabilities of pre-trained language models like RoBERTa-Large. ProMoT [39] is a two-stage LLM fine-tuning framework that first performs prompt tuning and then conducts fine-tuning, aiming to improve the out-of-domain generalization of in-context learning. Different from those approaches, our work aims to simultaneously improve ID and OOD generalization through a unified data and model-driven framework. In addition, we also enhance the discriminability of relevance modeling on fine-grained relevance degrees for better user experience, which may be overlooked by previous approaches.

## 3 PROBLEM FORMULATION

In this work, we consider the relevance modeling for search engines. Let $Q = \{q_1, q_2, \ldots, q_m\}$ and $R = \{r_1, r_2, \ldots, r_n\}$ denote the set of all queries and items respectively, where each query or item consists of a list of words describing its information content. Let $\mathcal{Y}$ denote the label space. We use a 3-point scale $y_i \in \mathcal{Y}$ to denote the graded relevance corresponding to the item $r_i$, i.e., strong relevance, weak relevance and irrelevance. Our training dataset, denoted as $\mathcal{D}_{id} = \{(q_i, r_i, y_i)\}_{i=1}^N$, is generated by sampling from the distribution $\mathcal{P}_{id}$. For simplicity, we will indistinctly use the notation $\mathcal{P}_{id}$ to refer to the joint and marginal distributions of $(Q, R, \mathcal{Y})$. A sample $x$ is called ID sample if $x \sim \mathcal{P}_{id}$. Let $\mathcal{P}_{ood} \neq \mathcal{P}_{id}$ denote a distribution which is 'far away' from $\mathcal{P}_{id}$. We refer to a sample $x \sim \mathcal{P}_{ood}$ as OOD sample.

In real-world search engines, a larger but unlabeled dataset $\mathcal{U} = \left\{\left(q'_i, r'_i\right)\right\}_{i=1}^U$ can be derived from search logs. We exploit data-level generalization improvements and select the most informative OOD samples for labeling to maximize the generalization performance. Let $\mathcal{D}_{ood}$ denote the resulting OOD samples that are selected from $\mathcal{U}$ and labeled for data enrichment. $\mathcal{D}_{ood}$ follows the distribution $\mathcal{P}_{ood}$. Based on $\mathcal{D}_{id}$ and $\mathcal{D}_{ood}$, the objective of relevance modeling involves learning a scoring function $f(\cdot, \cdot) : Q \times R \to \mathbb{R}$ that assesses the relevance degree $y_i$ for each query-item pair $(q_i, r_i)$, with the goal of minimizing the following generalization error:

$$\mathbb{E}_{(q,r,y)\sim\mathcal{P}_{id}}[\mathcal{L}(f(q,r),y)] + \mathbb{E}_{(q,r,y)\sim\mathcal{P}_{ood}}[\mathcal{L}(f(q,r),y)], \quad (1)$$

where $\mathcal{L}$ denotes the loss function.

## 4 METHODOLOGY

The overall systematic framework of the proposed DaRL is illustrated in Fig.1. It has three components: (i) *Distribution-Aware Sample Augmentation (DASA)* as described in Sec. 4.1, (ii) *Linear-Probing then Fine-Tuning (LPFT)* in Sec. 4.2, and (iii) *Over-Confidence Calibration (OCC)* in Sec. 4.3. The query-item pair is encapsulated within the pattern template *'Does item$_i$ satisfy the query $q_i$ search intent?'*, which is then processed by the LLM backbone to obtain the input representation. In contrast to the Pattern-Exploiting Training (PET) method [34], which aligns relevance label predictions with the generation of verbalizer tokens, we introduce a Multi-Layer Perceptron (MLP) layer, taking the model's hidden states as input, to perform relevance classification.

### 4.1 Distribution-Aware Sample Augmentation

We will leverage the vast amount of unlabeled data to enhance our model's out-of-distribution generalization capabilities. However, the unlabeled dataset comprises both in-distribution (ID) and out-of-distribution (OOD) samples. Annotating the entire dataset requires a significant amount of manual effort. Therefore, we design a novel distribution-aware strategy which utilizes OOD detection algorithms to identify OOD samples from the unlabeled dataset. These identified samples are subsequently manually annotated and utilized to enrich the training dataset.

In particular, we utilize the distance-based OOD detection method proposed by the authors in [19], which measures the distribution distance in the representation space between the unlabeled data and the original training set. We assume that the distribution of the training set follows a Gaussian distribution $\mathcal{P}_{id} = \mathcal{N}(\bar{x}, \Sigma)$, where $\bar{x}, \Sigma$ are estimated from the representation of the training data $\mathcal{D}_{id}$. The Mahalanobis distance is then used to measure the distance between each unlabeled sample and the training distribution:

$$d_{Mahal}(x^u, \mathcal{P}_{id}) = \sqrt{(g(x^u) - \bar{x})^T \Sigma^{-1} (g(x^u) - \bar{x})}, \quad (2)$$

where $x^u$ is the sample from the unlabeled dataset, $g(\cdot)$ maps samples to the representation space. If the Mahalanobis distance exceeds a certain threshold, we regard the sample as an OOD sample of the training set. Mahalanobis distance can be regarded as a global distance measure for OOD samples. Given $\bar{x}, \Sigma^{-1}$, let $b$ be the dimension of $g(\cdot)$, the time complexity of computing squared $d_{Mahal}(x^u, \mathcal{P}_{id})$ is $O(b^2)$.

Moreover, we introduce a local distance measure for samples [36], calculating the distance between any unlabeled sample and its nearest neighbor in the training set:

$$d_{knn}(x^u) = \min_{x_i \in \mathcal{D}_{id}} 1 - Cos(g(x^u), g(x_i)), \quad (3)$$

where $Cos(\cdot)$ denotes the cosine similarity. We use Faiss [17] for efficient nearest neighbor search. After obtaining the Mahalanobis distance and the nearest neighbor distance, we consider the samples where both metrics exceed specific thresholds as OOD samples:

$$\mathcal{U}_{ood} = \{x^u : d_{Mahal}(x^u, \mathcal{P}_{id}) > d_1 \land d_{knn}(x^u) > d_2\}. \quad (4)$$

We use the best ID F1 scores on the validation set to calculate the threshold $d_1$ and $d_2$ respectively. Then, we manually label $\mathcal{U}_{ood}$ and use it as augmented training data $\mathcal{D}_{ood}$ for model fine-tuning, mitigating the long-tail effect of the original training set on these out-of-distribution data. Therefore, the distribution-aware augmented training set $\mathcal{D}_{aug}$ is:

$$\mathcal{D}_{aug} = \mathcal{D}_{id} \cup \mathcal{D}_{ood}. \tag{5}$$

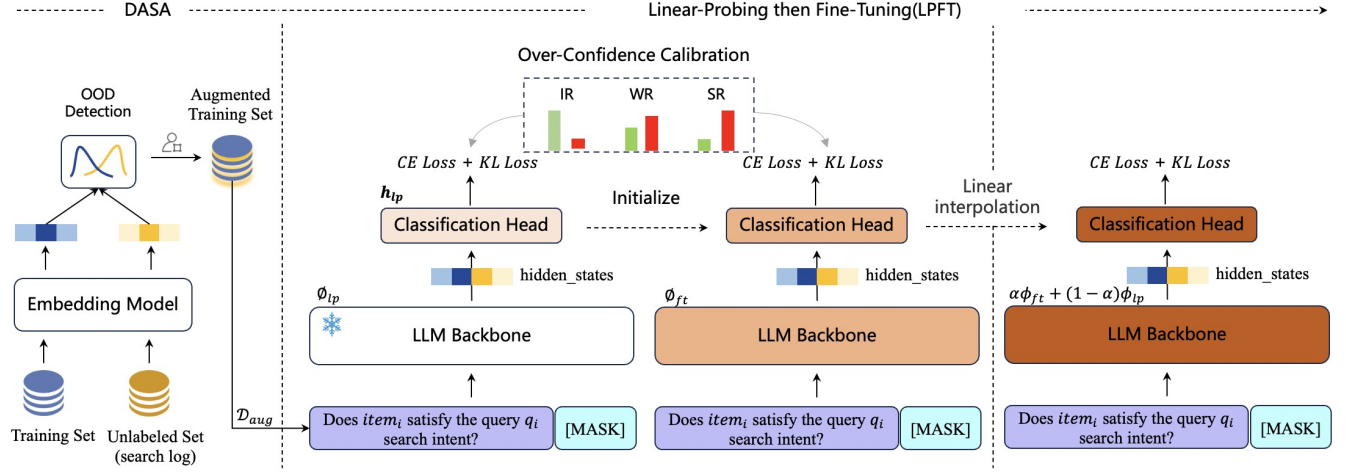

Figure 1: Overview of our proposed framework. DaRL contains three major components: (1) *Distribution-Aware Sample Augmentation (DASA)*, designed to leverage out-of-distribution detection for strategic sample enhancement, consequently improving the model's robustness when faced with data distribution drift; (2) a multi-stage training approach, dubbed *Linear-Probing then Fine-Tuning (LPFT)*, dedicated to reinforcing model robustness as it undergoes fine-tuning with the enriched OOD dataset; and (3) an auxiliary loss based on KL divergence, aimed at addressing issues of overconfidence, while simultaneously advancing the discriminative capabilities of LLM-based relevance models across diverse fine-grained degrees of query-item relevance.

More sophisticated strategies may be developed to select better distance measures for sample augmentation [22, 43], but it is out of scope of this paper.

## 4.2 Linear-Probing then Fine-Tuning

We fine-tune the relevance model with the OOD augmented dataset. However, the fine-tuning process may lead to catastrophic forgetting, resulting in a performance decline on in-distribution data. To strike a balance between the ID and OOD data, we employ the adapted Linear-Probing then Fine-Tuning (LPFT) method [46] to train the model.

LPFT is a multi-stage training method to reduce distortion during fine-tuning. In the first stage, adapted linear probing freezes the pre-trained LLM backbone parameters $\theta$ and only trains the classifier $h$ with a small set of additional parameters:

$$h_{lp} = \underset{h}{argmin} \mathcal{L}(\mathcal{D}_{aug}; \theta, h), \tag{6}$$

In implementation, we use the PEFT method LoRA [14] to train a good $h_{lp}$. After the first stage, we obtain the global model weights $\phi_{lp}$, including the pre-trained backbone parameters $\theta$ and the classifier header $h_{lp}$.

In the second stage, LPFT initializes the classifier with parameters from linear probing and then fine-tunes all parameters:

$$\phi_{ft} = \underset{\theta,h}{argmin} \mathcal{L}(\mathcal{D}_{aug}; \theta, h), \tag{7}$$

where $h$ is initialized by $h_{lp}$ from the first stage.

In the final stage, we perform a linear weight interpolation using the results from the previous two stages to enhance the overall performance of the model:

$$\phi = \alpha\phi_{ft} + (1-\alpha)\phi_{lp}, \tag{8}$$

where $\alpha$ is the interpolation coefficient. The adapted LPFT method can reduce the intensity of parameter changes in the LLM, which helps to mitigate the risk of catastrophic forgetting.

## 4.3 Over-Confidence Calibration

In our specific application context, the relevance score, obtained from $f(\cdot, \cdot)$, fulfills two crucial roles: firstly, it enables the categorization of items into various relevance grades according to predefined thresholds; secondly, it acts as a pivotal sorting criterion for the subsequent learn-to-rank layer. Consequently, ensuring that the relevance score accurately mirrors the level of similarity is imperative. Nonetheless, the conversion of relevance modeling to the generation of verbalizer tokens introduces the dilemma of over-confidence, as highlighted by Schick et al. [34]. This issue arises because the predicted probability distributions over the vocabulary tokens gravitate towards a single class, failing to genuinely represent the nuanced degrees of relevance similarity. As depicted in Figure 2, the model's predicted scores for weakly relevant samples predominantly cluster at both extremes of the [0,1] interval, resulting in a noticeable overlap with the predictions for samples deemed strongly relevant.

For the relevance model to aptly support subsequent phases, it's crucial that the predicted probability scores depict a spectrum of relevance degrees. Addressing this, we introduce a method to calibrate over-confidence in relevance models, aiming to enhance their performance and reliability.

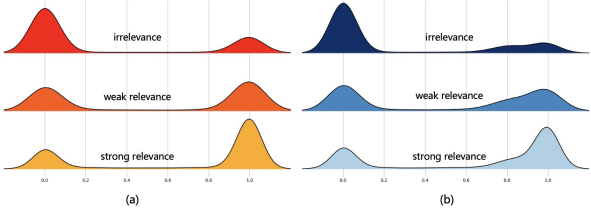

Figure 2: Comparing the effects of the Over-Confidence Calibration(OCC) module: (a) The score distribution for the test set pairs classified as strongly relevant (SR), weakly relevant (WR), and irrelevant (IR) without the inclusion of the OCC module. (b)The score distribution of the same categories with the inclusion of the OCC module.

In our setting, relevance is categorized into three distinct levels: strong relevance, weak relevance, and irrelevance. To model the relevance between a query-item pair, we employ a point-wise binary classification approach. We assume a prior distribution $q(x)$ over the labels (0, 1) as outlined below:

$$q(x) = \begin{cases} [1-\rho, \rho], & \text{if } x \text{ is IR} \\ [2\rho, 1-2*\rho], & \text{if } x \text{ is WR} \\ [\rho, 1-\rho], & \text{if } x \text{ is SR}, \end{cases} \quad (9)$$

where $\rho$ is a distribution smooth factor, with a default value of 0.1 in our setting. Please note that $\rho$ should be tuned according to the specific characteristics of the dataset or task at hand. In our case, the optimal parameter $\rho$ is determined through grid search.

An auxiliary loss $\mathcal{L}_{kl}$ based on KL divergence is introduced to adjust the distribution of the predicted values, and the overall loss is:

$$\mathcal{L} = \mathcal{L}_{ce} + \mathcal{L}_{kl}, \quad (10)$$

where $\mathcal{L}_{kl} = KL(p(x), q(x))$ is the KL divergence quantifies the discrepancy between the model's predicted score distribution $p(x)$ and the prior distribution $q(x)$. And $\mathcal{L}_{ce}$ is the cross-entropy loss for the binary classification task.

# 5 EXPERIMENTS
## 5.1 Experimental Setting

*5.1.1 Datasets.* To facilitate a better comparison of the performance of all models in our actual application scenarios, we select the real-world industry data of the Alipay insurance search and the statistical information of the dataset is provided in Table 1. The original training dataset $\mathcal{D}$ is annotated using crowdsourcing, where *SR*, *WR*, and *IR* represent *strong relevance*, *weak relevance*, and *irrelevance* labels, respectively. Additionally, we construct the distribution-aware augmented training set $\mathcal{D}_{aug}$ using the DASA pipeline introduced in this paper. The unlabeled dataset $\mathcal{U}$ is obtained by sampling from online search logs. Since our application scenario is insurance search, there is no domain-adapted dataset available, we use real business data to evaluate the proposed model.

Table 1: Statistical information of the experimental dataset. Please note that the quantity of items has been omitted for confidentiality purposes.

| Dataset | #Sample | #Query | #Item | #SR. | #WR. | #IR. |
|---|---|---|---|---|---|---|
| Train $\mathcal{D}_{id}$ | 199663 | 42672 | - | 45220 | 8582 | 145861 |
| Train $\mathcal{D}_{aug}$ | 226295 | 52187 | - | 47219 | 9028 | 170048 |
| Test $\mathcal{T}_{id}$ | 24958 | 15410 | - | 5605 | 1059 | 18294 |
| Test $\mathcal{T}_{ood}$ | 6465 | 3896 | - | 1152 | 1725 | 3588 |
| Test $\mathcal{T}_{prod}$ | 56882 | 17896 | - | 17561 | 13746 | 25575 |
| Unlabeled $\mathcal{U}$ | 2440883 | 341058 | - | - | - | - |

To provide a more comprehensive comparison of the effects of different methods, we use three test datasets, namely:

- **Test set $\mathcal{T}_{id}$** is the in-distribution test dataset with the same distribution as the original training set $\mathcal{D}_{id}$.
- **Test set $\mathcal{T}_{ood}$** is the out-of-distribution test dataset which is constructed according to the DASA module.
- **Test set $\mathcal{T}_{prod}$** is sourced from the production environment, comprising samples from $\mathcal{T}_{id}$ and $\mathcal{T}_{ood}$, along with long-tail samples from online search logs. The number of long-tail samples is roughly equivalent to $\mathcal{T}_{id}$.

*5.1.2 Baseline Models.*
- **DSSM** [15] is a deep learning model that aims to measure the semantic similarity between texts by mapping them into a low-dimensional semantic space.
- **BERT** is the classical pre-trained bidirectional transformers [5]. In this paper, **BERT-325M** is utilized as a fully interactive model to predict the relevance between queries and items, which is based on the *RoBERTa-wwm-ext-large*[2] version of the pre-trained model provided by the authors in [4].
- **AntGLM** builds upon **GLM** [7], tailored specifically to leverage Alipay's extensive datasets. The base version of **AntGLM** is refined through the method of Pattern-Exploiting Training (PET) for the relevance task [34].
- **ProMoT** [39] is a two-stage LLM fine-tuning method that first performs prompt tuning with frozen pretrained LLM parameters, and then conducts fine-tuning with the previously trained prompt being frozen.

*5.1.3 Evaluation Metrics.* We use F1 score (F1) and Accuracy (acc) to measure the performance of models, which are two commonly used metrics in evaluating classification models. Both higher metrics represent better model performance.

*5.1.4 Implementation Details.* We build the proposed DaRL based on AntGLM, and optimize it by Adam with a learning rate of 0.0005. The training batch size is 512 and the hidden size is 1024 for DaRL-0.3B (2048 for DaRL-2B, 4096 for DaRL-10B). The distribution-aware augmented training set $\mathcal{D}_{aug}$ is acquired according to Eq. 5 and $\mathcal{D}_{ood}$ is annotated by crowdsourcing. We also implement ProMoT based on AntGLM-0.3B and train ProMoT on $\mathcal{D}_{aug}$ using Adam and the same learning rate as DaRL.

[2]https://github.com/ymcui/Chinese-BERT-wwm

Table 2: Main results on our industrial data. The best performed methods in each metric are highlighted in bold. Significant improvements over the best baseline are marked with ∗ (t-test, $p < 0.05$)

| Method | Test $\mathcal{T}_{prod}$ | | Test $\mathcal{T}_{id}$ | | Test $\mathcal{T}_{ood}$ | |
|---|---|---|---|---|---|---|
| | F1 | acc | F1 | acc | F1 | acc |
| DSSM | 0.6034 | 0.5497 | 0.6372 | 0.5495 | 0.5370 | 0.5502 |
| BERT-325M | 0.6640 | 0.6770 | 0.7350 | 0.7037 | 0.5427 | 0.5404 |
| AntGLM-0.3B | 0.6995 | 0.7071 | 0.8908 | 0.8841 | 0.5498 | 0.5378 |
| ProMoT-0.3B | 0.7720 | 0.7751 | 0.9252 | 0.9251 | 0.6431 | 0.6431 |
| DaRL-0.3B | 0.7691 | 0.7787∗ | 0.9349 | 0.9345 | 0.6678∗ | 0.6818∗ |
| DaRL-2B | 0.7726 | 0.7792∗ | 0.9368 | 0.9365 | 0.6704∗ | 0.6843∗ |
| DaRL-10B | **0.7752∗** | **0.7802∗** | **0.9399∗** | **0.9395∗** | **0.6730∗** | **0.6864∗** |

### 5.2 Experimental Results

*5.2.1 Performance Comparison.* Table 2 provides a comprehensive comparison between DaRL and other baseline methods. The experimental results indicate that DSSM underperforms, as it is a basic two-tower model that only encodes the query and item separately. Further comparison with BERT-325M reveals that pre-trained models leveraging large corpora can offer additional benefits, with the interaction-based model demonstrating more advanced text relevance modeling abilities compared to representation-based models.

AntGLM-0.3B outperforms BERT-325M on the in-distribution test set $\mathcal{T}_{id}$, the out-of-distribution test set $\mathcal{T}_{ood}$, and the production test set $\mathcal{T}_{prod}$. This can be attributed to the fact that AntGLM-0.3B is customized to leverage Alipay in-house data. ProMoT-0.3B outperforms AntGLM-0.3B on all test sets, demonstrating superior performance due to its enhanced ability to generalize and the effective use of the augmented training set. Our proposed DaRL-0.3B outperforms ProMoT-0.3B on both the ID and OOD test sets in terms of F1 score, given the same model parameter scale. Furthermore, models with larger-scale parameters, such as DaRL-2B and DaRL-10B, demonstrate improved performance, indicating enhanced model capacities. However, models with larger parameters require substantial computational resources (i.e., GPUs, TPUs) and leads to hundreds of milliseconds delay, which are not feasible for industrial scenarios. Therefore, we have deployed the DaRL-0.3B model to provide online real-time serving services.

*5.2.2 Ablation Study.* In this section, we conduct ablation experiments to assess the efficacy of each component in our approach. The baseline model for the ablation study is AntGLM-0.3B with PET tuning [34]. Four sets of ablation experiments are conducted, including: (1) **Exp1** to assess the effectiveness of the classification head in relevance modeling; (2) **Exp2** to introduce the Over-Confidence Calibration (OCC) module; (3) **Exp3** to introduce the Distribution-Aware Sample Augmentation (DASA) module; and (4) **Exp4** to evaluate the effectiveness of the multi-stage training module-LPFT.

Throughout the ablation study presented in Table 3, the following observations are made:

Table 3: Ablation Study with different settings of DaRL.

| Method | Test $\mathcal{T}_{id}$ | | Test $\mathcal{T}_{ood}$ | |
|---|---|---|---|---|
| | F1 | acc | F1 | acc |
| AntGLM-0.3B | 0.8908 | 0.8841 | 0.5498 | 0.5378 |
| Exp1(+MLP) | 0.9052 | 0.9037 | 0.5536 | 0.5564 |
| Exp2(+MLP+OCC) | 0.9126 | 0.9141 | 0.5757 | 0.5618 |
| Exp3(+MLP+OCC+OOD) | 0.9197 | 0.9152 | 0.6362 | 0.6495 |
| Exp4(+All) i.e. DaRL | **0.9349** | **0.9345** | **0.6678** | **0.6818** |

- We first investigate the impact of the classification layer. As shown in Table 3 Exp1, compared to the approach of generating verbalizer tokens with PET, the classification task head exhibits general improvement on both in-distribution and out-of-distribution test sets.
- Exp2 in Table 3 demonstrates the effectiveness of the Over-Confidence Calibration (OCC) strategy, resulting in a 2.2pt increase in F1 score on the OOD test set. Fig.2 presents a comparison of the score distributions between the models with and without the OCC module. Incorporating the OCC module, the relevance scores of DaRL are more evenly distributed in the range of [0.5, 1.0], rather than being concentrated around 1.0 for pairs with weak relevance. Additionally, it shows a reduced overlap with pairs exhibiting strong relevance.
- Exp3 which involves training the model on $\mathcal{D}_{aug}$, shows a significant 6.05pt improvement on the OOD test set, while maintaining comparable performance on the ID test set.
- The effectiveness of the adapted LPFT tuning strategy is validated in Exp4. As compared to Exp3, the multi-stage tuning strategy in Exp4 makes more comprehensive use of the OOD augmented samples, thereby improving performance on both ID and OOD test sets.

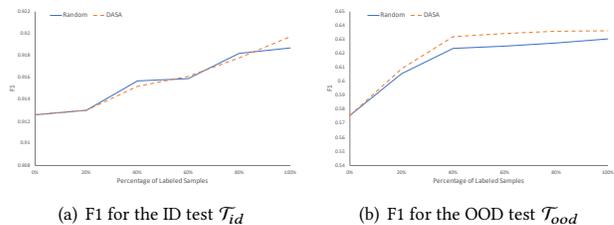

(a) F1 for the ID test $\mathcal{T}_{id}$  (b) F1 for the OOD test $\mathcal{T}_{ood}$

Figure 3: F1 comparison between the random sampling sample augmentation method (indicated by the blue solid line) and distribution-aware sample augmentation method (indicated by the orange dashed line) proposed by DASA. It is important to mention that the experiment is conducted under the same settings as Exp3 in Table 3, with the F1 results being reported for (a) the ID test $\mathcal{T}_{id}$ and (b) the OOD test $\mathcal{T}_{ood}$.

*5.2.3 Effectiveness of Distribution-aware Sample Augmentation Module.* We report the model's generalization performance with different proportions of $\mathcal{D}_{ood}$ samples in Fig.3. We have generated augmented training sets using two methods: random sampling sample augmentation and distribution-aware sample augmentation. We have then conducted a comparative analysis of the performance of models trained on the enhanced training sets, using identical settings. As shown in Fig.3, our findings indicate that both augmentation methods yield similar performance levels on the in-distribution test set, as measured by F1 results. However, on the out-of-distribution test set, the distribution-aware data augmentation method (DASA) demonstrates a notable performance enhancement. Our results suggest that increasing the proportion of OOD samples may further improve the model's performance.

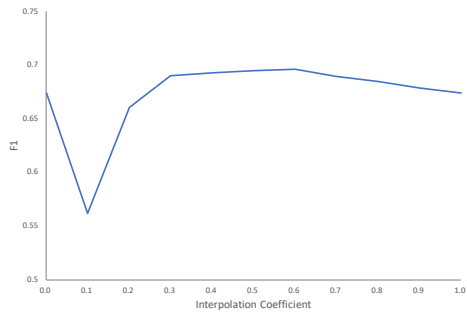

Figure 4: F1 results for the OOD test $\mathcal{T}_{ood}$ on the variation w.r.t. interpolation coefficient $\alpha$ in Eq.8 for the Linear-Probing then Fine-Tuning Module.

*5.2.4 Parameter Sensitivity of Linear-Probing then Fine-Tuning Module.* Based on the trends depicted in Fig. 4, it can be observed that: (1) The Linear-Probing then Fine-Tuning Module achieves improved results with an appropriate interpolation coefficient $\alpha$ in Eq. 8. (2) With $\alpha$ set to 0, only the classification head is fine-tuned, and with $\alpha$ set to 1, all model parameters are fine-tuned; however, neither approach leads to optimal performance. The optimal performance is attained at alpha = 0.6, necessitating a careful balance between linear probing and full parameter tuning.

## 5.3 Online A/B Testing

We deploy our proposed model on the Alipay search platform to provide online search service and conduct extensive AB experiments, with a baseline of the BERT model with equivalent size of parameters. At a confidence level of 95%, our model has improved the valid PV-CTR[3] by 1.2% and the PV-CVR[4] by 3.04% relative to the baseline model. Through manual evaluation, the irrelevant rate decreased by 3.98%, and the highly relevant rate increased by 3.17%. These experimental results in our industrial application scenario validate the effectiveness of our proposed approach.

The manual evaluation scheme uses a multi-person voting system and follows a "label-check-accept" process. We group and sample search queries and exposed documents based on online search logs, considering the search frequency over the past three months. Each sample is labeled by more than three trained annotators, using three categories: strongly relevant, weakly relevant, and irrelevant.

---

[3] the number of valid clicks divided by the number of searches
[4] the number of conversed search divided by the number of searches

The final evaluation sample set consists of approximately 20,000 to 40,000 samples.

## 6 CONCLUSION

This paper investigated the combination of distribution-aware data enrichment and robust fine-tuning strategies for LLM-based relevance modeling. We introduced a novel Distribution-Aware Robust Learning (DaRL) strategy for LLM-based relevance modeling, which utilized out-of-distribution detection techniques to improve the original training set and implemented a multi-stage fine-tuning strategy to enhance out-of-distribution generalizability while preserving in-distribution performance. The extensive experiments conducted on industrial data and online A/B testing confirmed the superiority of our approach and the effectiveness of its key components.

## REFERENCES

[1] Cory Barr, Rosie Jones, and Moira Regelson. 2008. The Linguistic Structure of English Web-Search Queries. In *2008 Conference on Empirical Methods in Natural Language Processing*. 1021–1030.
[2] Tom B. Brown, Benjamin Mann, Nick Ryder, Melanie Subbiah, Jared Kaplan, Prafulla Dhariwal, Arvind Neelakantan, Pranav Shyam, Girish Sastry, Amanda Askell, Sandhini Agarwal, Ariel Herbert-Voss, Gretchen Krueger, Tom Henighan, Rewon Child, Aditya Ramesh, Daniel M. Ziegler, Jeffrey Wu, Clemens Winter, Christopher Hesse, Mark Chen, Eric Sigler, Mateusz Litwin, Scott Gray, Benjamin Chess, Jack Clark, Christopher Berner, Sam McCandlish, Alec Radford, Ilya Sutskever, and Dario Amodei. 2020. Language Models are Few-Shot Learners. In *Advances in Neural Information Processing Systems 33*. 1877–1901.
[3] Zeyuan Chen, Wei Chen, Jia Xu, Zhongyi Liu, and Wei Zhang. 2023. Beyond Semantics: Learning a Behavior Augmented Relevance Model with Self-supervised Learning. In *Proceedings of the 32nd ACM International Conference on Information and Knowledge Management*. 4516–4522.
[4] Yiming Cui, Wanxiang Che, Ting Liu, Bing Qin, and Ziqing Yang. 2021. Pre-Training with Whole Word Masking for Chinese BERT. https://doi.org/10.1109/TASLP.2021.3124365
[5] Jacob Devlin, Ming-Wei Chang, Kenton Lee, and Kristina Toutanova. 2019. BERT: Pre-training of Deep Bidirectional Transformers for Language Understanding. In *Proceedings of the 2019 Conference of the North American Chapter of the Association for Computational Linguistics: Human Language Technologies*. 4171–4186.
[6] Mengnan Du, Fengxiang He, Na Zou, Dacheng Tao, and Xia Hu. 2024. Shortcut Learning of Large Language Models in Natural Language Understanding. *Commun. ACM* 67, 1 (2024), 110–120.
[7] Zhengxiao Du, Yujie Qian, Xiao Liu, Ming Ding, Jiezhong Qiu, Zhilin Yang, and Jie Tang. 2022. GLM: General Language Model Pretraining with Autoregressive Blank Infilling. In *Proceedings of the 60th Annual Meeting of the Association for Computational Linguistics*. 320–335.
[8] Guglielmo Faggioli, Laura Dietz, Charles L. A. Clarke, Gianluca Demartini, Matthias Hagen, and et al. 2023. Perspectives on Large Language Models for Relevance Judgment. In *Proceedings of the 2023 ACM SIGIR International Conference on Theory of Information Retrieval*. 39–50.
[9] Yixing Fan, Jiafeng Guo, Xinyu Ma, Ruqing Zhang, Yanyan Lan, and Xueqi Cheng. 2021. A Linguistic Study on Relevance Modeling in Information Retrieval. In *The Web Conference 2021*. 1053–1064.
[10] Jiafeng Guo, Yixing Fan, Qingyao Ai, and W. Bruce Croft. 2016. A Deep Relevance Matching Model for Ad-hoc Retrieval. In *Proceedings of the 25th ACM International Conference on Information and Knowledge Management*. 55–64.
[11] Jiafeng Guo, Yixing Fan, Qingyao Ai, and W. Bruce Croft. 2016. Semantic Matching by Non-Linear Word Transportation for Information Retrieval. In *Proceedings of the 25th ACM International Conference on Information and Knowledge Management*. 701–710.
[12] Junxian He, Chunting Zhou, Xuezhe Ma, Taylor Berg-Kirkpatrick, and Graham Neubig. 2022. Towards a Unified View of Parameter-Efficient Transfer Learning. In *The Tenth International Conference on Learning Representations*.
[13] Sebastian Hofstätter, Sheng-Chieh Lin, Jheng-Hong Yang, Jimmy Lin, and Allan Hanbury. 2021. Efficiently Teaching an Effective Dense Retriever with Balanced Topic Aware Sampling. In *Proceedings of the 44th International ACM SIGIR Conference on Research and Development in Information Retrieval*. 113–122.
[14] Edward J. Hu, Yelong Shen, Phillip Wallis, Zeyuan Allen-Zhu, Yuanzhi Li, Shean Wang, Lu Wang, and Weizhu Chen. 2022. LoRA: Low-Rank Adaptation of Large Language Models. In *The Tenth International Conference on Learning Representations*.

[15] Po-Sen Huang, Xiaodong He, Jianfeng Gao, Li Deng, Alex Acero, and Larry P. Heck. 2013. Learning deep structured semantic models for web search using clickthrough data. In *Proceedings of the 22nd ACM International Conference on Information and Knowledge Management*. 2333–2338.
[16] Shan Jiang, Yuening Hu, Changsung Kang, Tim Daly Jr., Dawei Yin, Yi Chang, and ChengXiang Zhai. 2016. Learning Query and Document Relevance from a Web-scale Click Graph. In *Proceedings of the 39th International ACM SIGIR conference on Research and Development in Information Retrieval*. 185–194.
[17] Jeff Johnson, Matthijs Douze, and Hervé Jégou. 2021. Billion-Scale Similarity Search with GPUs. *IEEE Transactions on Big Data* 7, 3 (2021), 535–547.
[18] Ananya Kumar, Aditi Raghunathan, Robbie Matthew Jones, Tengyu Ma, and Percy Liang. 2022. Fine-Tuning can Distort Pretrained Features and Underperform Out-of-Distribution. In *The Tenth International Conference on Learning Representations*.
[19] Kimin Lee, Kibok Lee, Honglak Lee, and Jinwoo Shin. 2018. A Simple Unified Framework for Detecting Out-of-Distribution Samples and Adversarial Attacks. In *Advances in Neural Information Processing Systems 31*. 7167–7177.
[20] Xiao Liu, Kaixuan Ji, Yicheng Fu, Weng Tam, Zhengxiao Du, Zhilin Yang, and Jie Tang. 2022. P-Tuning: Prompt Tuning Can Be Comparable to Fine-tuning Across Scales and Tasks. In *Proceedings of the 60th Annual Meeting of the Association for Computational Linguistics (Volume 2: Short Papers)*. 61–68.
[21] Yinhan Liu, Myle Ott, Naman Goyal, Jingfei Du, Mandar Joshi, Danqi Chen, Omer Levy, Mike Lewis, Luke Zettlemoyer, and Veselin Stoyanov. 2019. RoBERTa: A Robustly Optimized BERT Pretraining Approach. arXiv:1907.11692
[22] Haodong Lu, Dong Gong, Shuo Wang, Jason Xue, Lina Yao, and Kristen Moore. 2024. Learning with Mixture of Prototypes for Out-of-Distribution Detection. In *The Twelfth International Conference on Learning Representations*.
[23] Wenhao Lu, Jian Jiao, and Ruofei Zhang. 2020. TwinBERT: Distilling Knowledge to Twin-Structured Compressed BERT Models for Large-Scale Retrieval. In *Proceedings of the 29th ACM International Conference on Information and Knowledge Management*. 2645–2652.
[24] Xueguang Ma, Liang Wang, Nan Yang, Furu Wei, and Jimmy Lin. 2023. Fine-Tuning LLaMA for Multi-Stage Text Retrieval. arXiv:2310.08319
[25] Sean MacAvaney and Luca Soldaini. 2023. One-Shot Labeling for Automatic Relevance Estimation. In *Proceedings of the 46th International ACM SIGIR Conference on Research and Development in Information Retrieval*. 2230–2235.
[26] Humza Naveed, Asad Ullah Khan, Shi Qiu, Muhammad Saqib, Saeed Anwar, Muhammad Usman, Nick Barnes, and Ajmal Mian. 2023. A Comprehensive Overview of Large Language Models. arXiv:2307.06435
[27] OpenAI, Josh Achiam, Steven Adler, Sandhini Agarwal, Lama Ahmad, and et al. 2023. GPT-4 Technical Report. arXiv:2303.08774
[28] Long Ouyang, Jeffrey Wu, Xu Jiang, Diogo Almeida, Carroll L. Wainwright, and et al. 2022. Training language models to follow instructions with human feedback. In *Advances in Neural Information Processing Systems 35*. 27730–27744.
[29] Liang Pang, Yanyan Lan, Jiafeng Guo, Jun Xu, Shengxian Wan, and Xueqi Chen. 2016. Text Matching as Image Recognition. In *Proceedings of the Thirtieth AAAI Conference on Artificial Intelligence*. 2793–2799.
[30] Bo Peng, Xinyi Ling, Ziru Chen, Huan Sun, and Xia Ning. 2024. eCeLLM: Generalizing Large Language Models for E-commerce from Large-scale, High-quality Instruction Data. arXiv:2402.08831
[31] Nils Reimers and Iryna Gurevych. 2019. Sentence-BERT: Sentence Embeddings using Siamese BERT-Networks. In *Proceedings of the 2019 Conference on Empirical Methods in Natural Language Processing and the 9th International Joint Conference on Natural Language Processing*. 3980–3990.
[32] Stephen E. Robertson and Hugo Zaragoza. 2009. The Probabilistic Relevance Framework: BM25 and Beyond. *Foundations and Trends in Information Retrieval* 3, 4 (2009), 333–389.
[33] Keshav Santhanam, Omar Khattab, Jon Saad-Falcon, Christopher Potts, and Matei Zaharia. 2022. ColBERTv2: Effective and Efficient Retrieval via Lightweight Late Interaction. In *Proceedings of the 2022 Conference of the North American Chapter of the Association for Computational Linguistics: Human Language Technologies*. 3715–3734.
[34] Timo Schick and Hinrich Schütze. 2021. Exploiting Cloze-Questions for Few-Shot Text Classification and Natural Language Inference. In *Proceedings of the 16th Conference of the European Chapter of the Association for Computational Linguistics*. 255–269.
[35] Yelong Shen, Xiaodong He, Jianfeng Gao, Li Deng, and Grégoire Mesnil. 2014. Learning Semantic Representations Using Convolutional Neural Networks for Web Search. In *Proceedings of the 23rd International World Wide Web Conference, Companion Volume*. 373–374.
[36] Yiyou Sun, Yifei Ming, Xiaojin Zhu, and Yixuan Li. 2022. Out-of-Distribution Detection with Deep Nearest Neighbors. In *International Conference on Machine Learning*. PMLR, 20827–20840.
[37] Paul Thomas, Seth Spielman, Nick Craswell, and Bhaskar Mitra. 2024. Large Language Models can Accurately Predict Searcher Preferences. In *Proceedings of the 47th International ACM SIGIR Conference on Research and Development in Information Retrieval*. 1930–1940.
[38] Hugo Touvron, Thibaut Lavril, Gautier Izacard, Xavier Martinet, Marie-Anne Lachaux, Timothée Lacroix, Baptiste Rozière, Naman Goyal, Eric Hambro, Faisal Azhar, Aurélien Rodriguez, Armand Joulin, Edouard Grave, and Guillaume Lample. 2023. LLaMA: Open and Efficient Foundation Language Models. arXiv:2302.13971
[39] Yihan Wang, Si Si, Daliang Li, Michal Lukasik, Felix Yu, Cho-Jui Hsieh, Inderjit Dhillon, and Sanjiv Kumar. 2024. Two-stage LLM Fine-tuning with Less Specialization and More Generalization. In *The Twelfth International Conference on Learning Representations*.
[40] Miao Xiong, Zhiyuan Hu, Xinyang Lu, YIFEI LI, Jie Fu, Junxian He, and Bryan Hooi. 2024. Can LLMs Express Their Uncertainty? An Empirical Evaluation of Confidence Elicitation in LLMs. In *The Twelfth International Conference on Learning Representations*. https://openreview.net/forum?id=gjeQKFxFpZ
[41] Jun Xu, Xiangnan He, and Hang Li. 2020. Deep Learning for Matching in Search and Recommendation. *Foundations and Trends in Information Retrieval* 14, 2–3 (2020), 102–288.
[42] Haoran Yang, Yumeng Zhang, Jiaqi Xu, Hongyuan Lu, Pheng-Ann Heng, and Wai Lam. 2024. Unveiling the Generalization Power of Fine-Tuned Large Language Models. arXiv:2403.09162
[43] Jingkang Yang, Kaiyang Zhou, Yixuan Li, and Ziwei Liu. 2021. Generalized Out-of-Distribution Detection: A Survey. arXiv:2110.11334
[44] Linyi Yang, Yaoxian Song, Xuan Ren, Chenyang Lyu, Yidong Wang, Jingming Zhuo, Lingqiao Liu, Jindong Wang, Jennifer Foster, and Yue Zhang. 2023. Out-of-Distribution Generalization in Natural Language Processing: Past, Present, and Future. In *Proceedings of the 2023 Conference on Empirical Methods in Natural Language Processing*. 4533–4559.
[45] Linyi Yang, Shuibai Zhang, Libo Qin, Yafu Li, Yidong Wang, Hanmeng Liu, Jindong Wang, Xing Xie, and Yue Zhang. 2023. GLUE-X: Evaluating Natural Language Understanding Models from an Out-of-distribution Generalization Perspective. In *Findings of the Association for Computational Linguistics*. 12731–12750.
[46] Zhuoyi Yang, Ming Ding, Yanhui Guo, Qingsong Lv, and Jie Tang. 2022. Parameter-Efficient Tuning Makes a Good Classification Head. In *Proceedings of the 2022 Conference on Empirical Methods in Natural Language Processing*. 7576–7586.
[47] Wen Zan, Yaopeng Han, Xiaotian Jiang, Yao Xiao, Yang Yang, Dayao Chen, and Sheng Chen. 2023. SPM: Structured Pretraining and Matching Architectures for Relevance Modeling in Meituan Search. In *Proceedings of the 32nd ACM International Conference on Information and Knowledge Management*. 4923–4929.